\documentclass[10pt,letterpaper,twocolumn]{article} 

\usepackage{times,mathptmx}
\usepackage{ol2}
\usepackage{amsmath,amssymb}

\begin{document}

\twocolumn[ 

\title{Modeling of octave-spanning Kerr frequency combs using a generalized mean-field Lugiato-Lefever model}


\author{St\'ephane Coen,$^{1,*}$ Hamish G. Randle,$^1$ Thibaut Sylvestre,$^2$ and Miro Erkintalo$^1$}

\address{
$^1$ Physics Department, The University of Auckland, Private Bag 92019, Auckland, New Zealand\\
$^2$ Institut FEMTO-ST, D\'epartement d'Optique P. M. Duffieux, Universit\'e de Franche-Comt\'e, CNRS, Besan\c{c}on, France\\
$^*$Corresponding author: s.coen@auckland.ac.nz}

\begin{abstract}A generalized Lugiato-Lefever equation is numerically solved with a Newton-Raphson
  method to model Kerr frequency combs. We obtain excellent agreement with past experiments, even for an
  octave-spanning comb. Simulations are much faster than with any other technique despite including more modes
  than ever before. Our study reveals that Kerr combs are associated with temporal cavity solitons and dispersive
  waves, and opens up new avenues for the understanding of Kerr comb formation.
\end{abstract}

\ocis{230.5750, 190.5530, 190.4380}

 ] 

\noindent First observed in 2007, frequency comb generation in monolithic microring resonators has aroused
significant research interest \cite{kippenberg_microresonator-based_2011, ferdous_spectral_2011}. A minuscule
footprint, power efficiency, and CMOS-compatibility make said structures ideal for on-chip frequency comb
generation. Applications range from spectroscopy to telecommunications.

Although comb generation in high-Q Kerr resonators has been extensively reported experimentally, theoretical studies
are comparatively scarce. In part this deficiency can be linked to the intractable computational complexity of the
existing models. Indeed, the use of a nonlinear Schr\"odinger  (NLS) equation and appropriate coupling-mediated
boundary conditions requires hundreds of millions of roundtrips for convergence, owing to the exceedingly high Q of
the structures~\cite{agha_theoretical_2009}. Likewise the number of four-wave mixing-mediated coupling terms in
coupled mode equation models scales cubically with the number of modes, limiting such modeling to narrowband
combs~\cite{chembo_modal_2010}. Matsko \textit{et al.}~\cite{matsko_mode-locked_2011} also considered an analytic
approach but it is restricted to resonators of infinite intrinsic Q-factor and with group-velocity dispersion (GVD)
limited to 2nd order. Clearly, a growing demand exists for realistic yet computationally efficient methods for the
modeling of frequency combs in high-Q resonators.

In this Letter we report on direct numerical modeling of Kerr frequency combs using a generalized Lugiato-Lefever
(LL) equation~\cite{lugiato_spatial_1987}, and find good agreement with reported experiments. Significantly, the
conducted computations are far less intensive than previous methods, allowing for the rapid modeling of
octave-spanning combs with arbitrarily low repetition rates on a consumer grade computer. We also show that the
results obtained from the proposed model provide significant insights into Kerr combs. In particular, we highlight
how localized dissipative structures known as temporal cavity solitons (CSs)~\cite{leo_temporal_2010} are stable
stationary solutions of the governing equation, and how specific comb features can be intuitively described in terms
of CS propagation. We expect the reported technique to become an invaluable tool for the understanding and tailoring
of nonlinear optical phenomena in high-Q resonators.

\begin{figure}[b]
  \includegraphics[width = \columnwidth, clip = true]{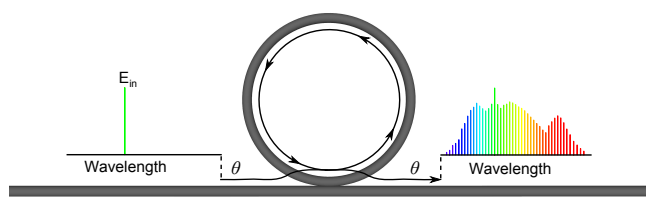}
  \vskip-2mm
  \caption{\small{(Color online) Schematic of the resonator configuration.}}
  \label{schematic}
\end{figure}
We consider a  typical ring resonator configuration (Fig.~\ref{schematic}): a continuous-wave (cw) driving field
$E_\mathrm{in}$ with power $P_\mathrm{in}=|E_\mathrm{in}|^2$ is coherently added to the lightwave circulating in the
resonator through a coupler with power transmission coefficient $\theta$. The fourth port of the coupler is used to
extract the output field. Mathematically the intracavity field $E^{(m+1)}(0,\tau)$ at the beginning of the
$(m+1)^\mathrm{th}$ roundtrip can be related to the field $E^{(m)}(L,\tau)$ at the end of the $m^\mathrm{th}$
roundtrip as:
\begin{equation}
  \label{boundary}
  E^{(m+1)}(0,\tau) = \sqrt{\theta}\,E_\mathrm{in}+\sqrt{1-\theta}\,E^{(m)}(L,\tau)e^{i\phi_0},
\end{equation}
where $\tau$ is the time, $L$ the roundtrip-length of the resonator, and $\phi_0$ the linear phase accumulated by
the intracavity field with respect to the the pump field over one roundtrip.

Assuming that light propagates in a single spatial mode, the evolution of the field through one roundtrip of the
resonator is governed by the well-known (generalized) NLS equation:
\begin{equation}
\label{GNLSE}
\frac{\partial E(z,\tau)}{\partial z} = -\frac{\alpha_\mathrm{i}}{2}E
    + i \sum_{k\geq 2}\frac{\beta_k}{k!}\left(i\frac{\partial}{\partial \tau }\right)^k E  +  i\gamma |E|^2 E\,.
\end{equation}
Here $\alpha_\mathrm{i}$ is the linear absorption coefficient inside the resonator, $\beta_k = d^k\beta/d\omega^k
|_{\omega = \omega_0}$ the dispersion coefficients associated with the Taylor series expansion of the propagation
constant $\beta(\omega)$ about the center frequency $\omega_0$ of the driving field, and $\gamma =
n_2\omega_0/(cA_\mathrm{eff})$ the nonlinearity coefficient with $n_2$ the nonlinear refractive index and
$A_\mathrm{eff}$ the effective modal area of the resonator mode.

The boundary conditions Eq.~\eqref{boundary} combined with the NLS Eq.~\eqref{GNLSE} form an infinite-dimensional
map that describes completely the dynamics of a ring resonator of any size or shape (toroid, racetrack, $\ldots$).
In low loss structures, the field varies only slightly between consecutive roundtrips, making direct simulations of
these equations very slow~\cite{agha_theoretical_2009}. In these conditions, however, it is well known that Eqs.
(\ref{boundary})--(\ref{GNLSE}) can be averaged into an externally driven NLS equation~(see, e.g.,
\cite{haelterman_dissipative_1992}),
\begin{align}
  \label{LL}
  t_\mathrm{R}\frac{\partial E(t,\tau)}{\partial t} =& \bigg[ -\alpha - i \delta_0 +iL\sum_{k\geq 2}\frac{\beta_k}{k!}\left(i\frac{\partial}{\partial \tau}\right)^k \nonumber \\
  & + i\gamma L |E|^2 \bigg] E + \sqrt{\theta}\,E_\mathrm{in},
\end{align}
where $t_\mathrm{R}$ is the roundtrip time, $\alpha = (\alpha_\mathrm{i}+\theta)/2$ describes the total cavity
losses, and $\delta_0 = 2\pi l - \phi_0$ is the detuning with $l$ the order of the cavity resonance closest to the
driving field. The continuous variable $t$ measures the \textit{slow time} of the cavity, and can be linked to the
roundtrip index as $E(t = mt_R,\tau) = E^{(m)}(0,\tau)$. Equation~\eqref{LL} coincides with the master equation
of~\cite{matsko_mode-locked_2011} and, with $\beta_k = 0$ for $k\geq 3$, is formally identical to the mean-field LL
model of a diffractive cavity \cite{lugiato_spatial_1987, scroggie_pattern_1994, lugiato_introduction_2003}. It has
also been extensively used for the description of passive cavities constructed of single-mode fibers
\cite{haelterman_dissipative_1992, coen_competition_1999, coen_continuous-wave_2001, leo_temporal_2010}. In
particular, spatial pattern formation and the so-called modulation instability (MI) studied in some of these earlier
works can be directly connected to frequency comb generation~\cite{lugiato_spatial_1987, scroggie_pattern_1994,
coen_continuous-wave_2001}. MI was also shown to occur in the normal GVD regime in presence of cavity boundary
conditions~\cite{haelterman_dissipative_1992, coen_competition_1999}, which is directly relevant to combs. We must
also note that the expressions of the characteristic bistable response of the LL model as well as that of the
intracavity MI gain \cite{haelterman_dissipative_1992} in fact coincide after normalization with corresponding
results obtained from the coupled-mode equations of \cite{chembo_modal_2010}, suggesting an intrinsic link between
the two approaches.

The generalized LL Eq.~\eqref{LL} holds in the limit of high finesse cavities $\mathcal{F} \gg 1$. For typical
high-Q resonators, the finesse $\mathcal{F}\sim 10^2$--$10^5$. Also, dispersion must be ``weak'' over one roundtrip,
$\sum_{k\geq 2} \beta_k L \Delta\omega^k/k! \lesssim \pi$, where $\Delta\omega$ is the (angular) spectral width of
the generated comb. This condition was found to be satisfied \textit{a posteriori} for all the results discussed
below, thereby asserting the validity of Eq. \eqref{LL} for the description of Kerr combs. The LL equation has two
substantial advantages compared to the infinite dimensional map Eqs. \eqref{boundary}--\eqref{GNLSE}. On the one
hand, Eq. \eqref{LL} can be numerically integrated with the split-step Fourier method using an integration step
corresponding to \textit{multiple roundtrips}, significantly reducing the computational burden in obtaining steady
state solutions. On the other hand, the steady state solutions can be obtained directly by setting $\partial
E/\partial t = 0$ and looking for the roots of the right-hand-side of Eq.~\eqref{LL}. Although the latter approach
does not yield insights into the dynamics of comb formation, it is computationally orders of magnitude more
efficient than split-step integration. Here we restrict our attention to stationary solutions obtained by a
multidimensional root-finding Newton-Rhapson method. Derivatives are computed with Fourier transforms and the span
of the temporal window coincides with $t_\mathrm{R}$, meaning the samples of our frequency grid are spaced by the
free-spectral range, $\mathrm{FSR} = 1/t_\mathrm{R}$.

\begin{figure}[t]
  \includegraphics[width = \columnwidth,clip=true]{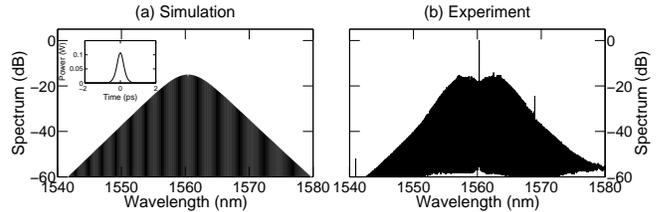}
  \vskip-2mm
  \caption{\small{(a) Steady state solution of Eq. \eqref{LL} for a critically-coupled,
    3.8 mm diameter MgF$_2$ whispering gallery mode resonator with a $40$ $\mu$m mode-field diameter and a loaded
    $Q = 1.90 \cdot 10^{9}$. $\mathrm{FSR}=18.2$~GHz;
    $\gamma = 0.032\ \mathrm{W^{-1} km^{-1}}$; $\beta_2 = -13\ \mathrm{ps^2\,km^{-1}}$; $\alpha = \theta = 1.75\cdot10^{-5}$;
    $P_\mathrm{in} = 55.6$ mW; $L= 11.9$ mm; $\delta_0 = 0.0012$. (b) Corresponding experimental spectrum
    after \cite{grudinin_frequency_2012}.}}
  \label{exmp1}
  \vskip-5mm
\end{figure}

As a first example, we plot in Fig. \ref{exmp1}(a) the \emph{intracavity} spectrum of a stable steady-state solution
of Eq. \eqref{LL} obtained with simulation parameters listed in the caption and approximating the experimental
values of \cite{grudinin_frequency_2012}. We note that some of these parameters have large uncertainties but this
does not invalidate our conclusions. Figure \ref{exmp1}(b) is the corresponding experimental \emph{output} spectrum,
and clearly excellent agreement is observed (abstraction must be made of the pump mode intensity which is modified
at the output by the reflected pump). Some of the discrepancies could be traced to effects unaccounted for in Eq.
\eqref{LL} such as wavelength-dependent losses or overlap integrals, or to experimental fluctuations. We also show
the temporal profile of the solution as an inset of Fig. \ref{exmp1}(a) so as to highlight that the intracavity
field corresponds to a $\sim 400$~fs-duration pulse with a peak power of 100~mW atop a weak cw background. It is
well-known that the LL equation possesses such localized-CS solutions repeating at the cavity repetition rate thus
forming a frequency comb in the spectral domain \cite{lugiato_introduction_2003,leo_temporal_2010}. In fact, in
solving Eq. \eqref{LL}, we have not found any type of stable steady-state comb solutions not made up of single or
multiple CSs. This strongly suggests that stable frequency combs generated in high-Q cavities generally correspond
to trains of CSs which is in good accordance with recent studies~\cite{ferdous_spectral_2011,
matsko_mode-locked_2011, matsko_hard_2012} and was already suggested in \cite{leo_temporal_2010}. Our analysis also
often reveals coexisting unstable states associated with MI and breathing CSs which can in practice preclude the
observation of stable combs. Care must therefore be taken in interpreting some reported experimental results: They
may be associated with rapid fluctuations and only appear stable because of the averaging of spectrum analyzers.

\begin{figure}[t]
\includegraphics[width = \columnwidth, clip = true]{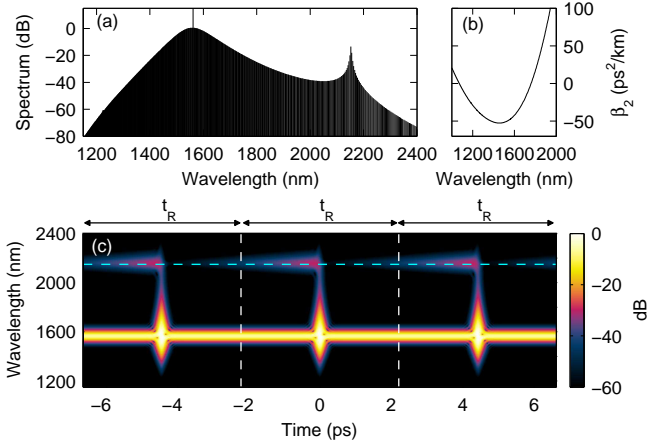}
  \vskip-2mm
  \caption{\small{(Color online) (a) Steady-state solution of Eq. \eqref{LL} with parameters mimicking a critically-coupled,
  $200\ \mu$m diameter silicon nitride resonator with a loaded $Q = 3\cdot 10^{5}$ approximated from
  \cite{okawachi_octave-spanning_2011}. Dispersion as per (b); $\mathrm{FSR} = 226$~GHz;
  $\gamma =  1\ \mathrm{W^{-1} m^{-1}}$; $\alpha = \theta = 0.009$; $P_\mathrm{in} = 755$ mW; $L=628\ \mu$m;
  $\delta_0 = 0.0534$. (c) Time-frequency representation of the simulation result calculated using a 100~fs gate
  function. Subsequent roundtrips are separated by vertical lines whilst the horizontal line
  indicates the predicted \v{C}erenkov wavelength.}}
  \label{exmp2}
  \vskip-5mm
\end{figure}

Our next example (Fig.~\ref{exmp2}) is an octave-spanning comb, strongly influenced by higher-order dispersion. The
simulation parameters are radically different and are approximated from the experiment of
\cite{okawachi_octave-spanning_2011} (see caption). Again, based on the same model Eq.~\eqref{LL}, we observe an
excellent agreement between the stable steady-state numerical spectrum plotted in Fig. \ref{exmp2}(a) and the
experimental measurements (see Fig.~2 in \cite{okawachi_octave-spanning_2011}). At this point we must emphasize
that, despite the large number of spectral modes included (1024), the computation time in obtaining the results
shown both in Figs. \ref{exmp1}(a) and \ref{exmp2}(a) was of the order of seconds on a standard computer. It is
precisely this unprecedented speed that is the key advantage of our technique. In fact, to our knowledge, the Kerr
comb of Fig.~\ref{exmp2}(a) has the largest number of optical modes obtained from a theoretical model. Simulating
octave-spanning combs with sub-40~GHz repetiton rates would necessitate to ramp up the number of modes to~4096. In
this case, computation time increases to about 2--3 minutes (or less if considering neighboring solutions), but is
still very reasonable.

A particular feature seen in Fig. \ref{exmp2}(a) is the strong narrowband low-frequency component centered about
2150 nm. A similar feature can be witnessed in the corresponding experimental measurements
\cite{okawachi_octave-spanning_2011}, and also in other previous experiments (see, e.g., Fig.~1 in
\cite{foster_silicon-based_2011}). We interpret this component as a \v{C}erenkov-like resonant dispersive wave (DW)
emitted by the CS circulating in the resonator \cite{dudley_supercontinuum_2006}. To show this explicitly, we plot
in Fig.~\ref{exmp2}(c) the time-frequency representation of the intracavity electric field. Here we exploit the
periodic boundary conditions, and expand the fast temporal axis across three cavity roundtrips. We can clearly see
how the intracavity field consists of a train of pulses atop a cw background (i.e., CSs) and we can identify the
narrowband 2150~nm component as DWs emitted by individual CSs. This is further highlighted by the dashed horizontal
line in Fig. \ref{exmp2}(c) which indicates the predicted wavelength $\lambda_\mathrm{DW}$ of a \v{C}erenkov-wave.
Specifically, neglecting the nonlinear contribution, the pertinent phase-matching condition governing the resonant
DW is $\beta(\omega_\mathrm{DW}) = \beta(\omega_\mathrm{CS})
-\beta_1(\omega_\mathrm{CS})\cdot(\omega_\mathrm{CS}-\omega_\mathrm{DW})$ \cite{dudley_supercontinuum_2006}, where
$\omega_\mathrm{DW}=2\pi c/\lambda_\mathrm{DW}$ and $\omega_\mathrm{CS}$ are the central (angular) frequencies of
the DW and the CS, respectively. With our numerical parameters, this condition yields $\lambda_\mathrm{DW} = 2149$
nm, in excellent agreement with the observed spectral peak in Fig. \ref{exmp2}(a). Because a (cavity) soliton in the
anomalous GVD regime can only excite DWs in the normal GVD regime, our observations suggest that the ubiquitous
asymmetry of Kerr combs towards the normal GVD regime may in fact be explained by the excitation of resonant DWs by
CSs in the anomalous GVD region in a way akin to supercontinuum generation~\cite{dudley_supercontinuum_2006}.
Finally, we note that DW emission has recently been described in terms of cascaded four-wave mixing, which further
highlights its relevance to Kerr combs~\cite{erkintalo_cascaded_2012}.

In conclusion, we have used a generalized LL equation to model high-Q resonator Kerr frequency combs. It can be
solved with a Newton-Raphson solver, providing results in a matter of seconds, much faster than any other technique,
and for widely different cases, all the way to octave-spanning combs with hundreds of modes. Our results are in good
agreement with experiments, proving that the LL model captures the essential physics of Kerr combs. We also find
that Kerr frequency combs are associated in the temporal domain with CSs~\cite{leo_temporal_2010} and associated
DWs. CSs are well known localized dissipative structures which have been studied extensively in the past, in
particular in the spatial domain \cite{lugiato_introduction_2003, scroggie_pattern_1994}, so we anticipate that the
LL model will provide both a deeper understanding to Kerr frequency combs and computational efficiency. The Newton
method also provides information about dynamical instabilities of Kerr combs through an eigenvalue analysis of the
Jacobian of the system, which is obtained at no extra cost. Some of our preliminary results highlight the presence
of Hopf bifurcations associated with breathing CSs, i.e., combs with periodically modulated features.

We thank Dr I. S. Grudinin for kindly providing experimental data. We also acknowledge support from the Marsden Fund
of The Royal Society of New Zeland.

\clearpage

\renewcommand\refname{References with titles}

\end{document}